# Bimodal Cochlear Implants: Measurement of the Localization Performance as a Function of Device Latency Difference

*Rebecca C. Felsheim[*+,1,2], Sabine Hochmuth[*,3], Alina Kleinow[1], Andreas Radeloff[2,3,4], Mathias Dietz[1,2]*

[1] *Department für Medizinische Physik und Akustik, Carl von Ossietzky Universität Oldenburg, Oldenburg, Germany*

[2] *Cluster of Excellence "Hearing4All", Oldenburg, Germany*

[3] *Department of Otolaryngology, Head and Neck Surgery, Carl von Ossietzky Universität Oldenburg, Oldenburg, Germany*

[4] *Research Center Neurosensory Science, University of Oldenburg, Carl von Ossietzky Universität Oldenburg, Oldenburg, Germany*

*\* joint first authors*

[+] *corresponding author: rebecca.felsheim @uni-oldenburg.de*

## Abstract

Bimodal cochlear implant users show poor localization performance. One reason for this is a difference in the processing latency between the hearing aid and the cochlear implant side. It has been shown that reducing this latency difference acutely improves the localization performance of bimodal cochlear implant users. However, due to the frequency dependency of both the device latencies and the acoustic hearing ear, current frequency-independent latency adjustments cannot fully compensate for the differences, leaving open which latency adjustment is best. We therefore measured the localization performance of eleven bimodal cochlear implant users for multiple cochlear implant latencies. We confirm previous studies that adjusting the interaural latency improves localization in all eleven bimodal cochlear implant users. However, the latency that leads to the best localization performance for most subjects was not at the latency calculated to compensate for the interaural difference at intermediate frequencies (1 kHz). Nine of eleven subjects localized best with a cochlear implant latency that was slightly shorter than the calculated latency compensation.

# Introduction

Binaural hearing, i.e., hearing with two ears, enables listeners to pinpoint the direction of a sound with remarkable accuracy. Trained normal hearing listeners can distinguish directions of two sources separated by angles as small as 1° in front of the head (Mills, 1958). In listeners with impaired hearing, the ability to localize sounds is reduced compared to normal hearing listeners. In a study by Dorman et al. (2016) which compared the localization performance of different subject groups, bilateral hearing aid users had an average root mean square error (RMSE) of 12° and bilateral cochlear implant (CI) users had an average RMSE of 29°. This stands against an average RMSE of 5° in an approximately age-matched normal hearing control group. An even larger average RMSE of 62°, close to chance performance, was reported in a small group of bimodal CI users with a hearing aid contralateral to the CI.

Most bimodal CI users used to be severely hearing impaired in both ears but were only eligible to receive one CI due to funding restrictions. With a high-power hearing aid, these patients receive some low-frequency hearing to complement their dominant CI, but, not surprisingly, they perform very poorly in sound localization, with little to no spectral overlap between CI and hearing aid stimulation. The bimodal participants in Dorman et al. (2016) appear to fall into this category. However, when funding is less restrictive, a second very different group of bimodal CI users can be found: Those who have only a moderate hearing loss in one ear, that is best treated with a hearing aid, and a severe to profound hearing loss in the other ear, treated with a CI. These patients can already understand speech with their hearing aid alone and opt for the CI to get a more symmetric hearing sensation as well as to restore their spatial hearing. While better spatial hearing should be possible in this group due to a spectral overlap of both stimulation modalities, even this group has a rather poor sound localization with an average RMSE of 53° (Angermeier et al., 2021).

Multiple reasons are conceivable for the lower localization performance of the bimodal CI users. For instance, the human brain may not be able to integrate the two very different modalities of electrical and acoustic hearing well enough to enable binaural processing. Other possible causes are the mismatches between the two ears caused by the two different devices and differently impaired ears. These include tonotopic mismatch, level mismatch, latency mismatch, and a mismatch in the spectral content between the two ears, caused by the CI not being able to selectively stimulate the apical turn and high-frequency limitation in the acoustic hearing ear (see Pieper et al., 2022 for a review).

The level mismatch has its roots in the often-independent fitting of the two devices and the reduced dynamic range especially in electric hearing. While it might be possible to balance the levels in the two devices for a specific stimulus, the reduced dynamic range makes it difficult to balance the levels for a wider range of stimuli.

The last dimension of mismatch is latency: In the acoustically stimulated ear, this latency consists first of the frequency and manufacturer-dependent processing latency of the hearing aid (1 - 7 ms; Zirn et al., 2015) and on the frequency and level-dependent latency of the human ear (1 - 8 ms; Ruggero & Temchin, 2007). For the CI side, only the latency of the device needs to be considered, which is frequency dependent and ranges from 0.5 - 7 ms for MED-EL CIs (Zirn et al., 2015), simulating the frequency dependency of the human ear, and a nearly frequency independent latency of about 12 ms for Cochlear devices (Engler et al., 2020). These different latencies add up to a latency mismatch of several milliseconds, vastly exceeding even the highest acoustically caused interaural time differences (ITDs) of about 740 µs (Hartmann & Macaulay, 2014), corresponding to a sound from 90°.

For bimodal MED-EL CI users, the latency on the CI side is lower than the combined latency (hearing aid latency + ear latency) on the acoustic side for all frequencies and hearing aid manufactures. Therefore, adding an additional delay to the CI processor is a promising option that is included in the most recent versions of the clinical fitting software Maestro since 2020. At present, only a frequency independent delay can be added. When ear and device latencies are known or estimated, the presumed interaural latency mismatch can be derived. We call it "estimated latency mismatch", and this value can be added to the CI processor.

Angermeier et al. (2021) measured the localization performance in bimodal MED-EL CI users with and without compensating for the estimated latency mismatch. They found that compensating for the estimated latency mismatch in bimodal patients acutely improved the localization performance significantly from an average RMSE of 53° to 38°. No further improvement was found after a familiarization period of three weeks Angermeier et al. (2023).

As the MED-EL CI latency is very similar to the latency of an acoustically stimulated ear, the estimated latency mismatch is approximately equal to the hearing aid latency, especially at frequencies near 1 kHz. At lower frequencies, the CI is a bit slower, and at higher frequencies even a bit faster (Zirn et al., 2015). Angermeier et al. (2021) therefore tested three different compensation delays: One that is equal to the hearing aid latency, 1 ms more, and 1 ms less. For 5 of 7 listeners, the lowest RMSE was obtained with the largest delay. Localization bias without added CI latency was always to the side of the CI and decreased as expected with increasing CI delay. But even for the largest delay, 6 of 7 listeners had a residual bias towards the CI. This may have been caused by mismatches in other dimensions, such as level (Pieper et al., 2022). We hypothesize that an even larger CI delay would compensate for the residual bias, possibly even further reducing the RMSE as well. A good compensation could possibly be able to reduce the RMSE to the range of bilateral CI users, as also hypothesized by Pieper et al. (2022). In this work, we have therefore measured the localization performance of eleven bimodal CI users with MED-EL CIs over a wider range of added CI latencies.

## Methods

### Estimated Latency Difference

Zirn et al. (2015) estimated the latency difference between the two ears based on measurements of the wave V of the auditory brainstem response. Following their approach, the latency on the acoustic hearing side is given as

$$\tau_a(f) = \tau_{HA}(f) + \tau_{ear}(f) + \tau_V \quad (1)$$

with $\tau_{HA}(f)$ being the potentially frequency-dependent latency of the hearing aid, $\tau_{ear}(f)$ the frequency dependent-latency of the ear and $\tau_V$ the frequency-independent latency of the neural processing until wave V occurs. Similarly, for the electrically hearing side, the latency can be obtained as

$$\tau_e(f) = \tau_{CI}(f) + \tau_{add} + \tau_V \quad (2)$$

where $\tau_{CI}(f)$ is the frequency-dependent latency of the CI and $\tau_{add}$ an additional latency that can be added to the processing latency. The value $\tau_V$ is assumed to be the same for the acoustic and electric hearing side. The latency difference is then given as

$$\Delta\tau(f) = \tau_a(f) - \tau_e(f) = (\tau_{HA}(f) + \tau_{ear}(f)) - (\tau_{CI}(f) + \tau_{add}) \quad (3)$$

For MED-EL devices the difference between $\tau_{ear}$ and $\tau_{CI}$ is very small. At 500 Hz $\tau_{ear}$ is marginally smaller, at 1 kHz they are virtually identical and at 2 and 4 kHz $\tau_{ear}$ is even a bit larger (Zirn et al., 2015). However, currently the clinical MED-EL software only allows for a frequency-independent additional latency. Therefore, a frequency-independent version of $\Delta\tau(f)$ needs to be estimated. We follow the approach of Angermeier et al. (2021) and assume an average equality between $\tau_{ear}$ and $\tau_{CI}$. This simplifies (3) to

$$\Delta\hat{\tau} = \hat{\tau}_{HA} - \tau_{add} \quad (4)$$

as frequency-independent estimation of the latency difference between the two ears, with the frequency-independent estimate of the hearing aid latency $\hat{\tau}_{HA}$.

For most participants, the hearing aid latencies ($\hat{\tau}_{HA}$) were measured with the hearing aid analyzer unit ACAM 5 (Acousticon GmbH, Reinheim, Germany) using tone bursts in the frequency range of 500 to 4000 Hz. The hearing aid latency was measured from the time difference of the rising flank of the envelope between input and output. The median of the measured latencies in the frequency range of 500 to 4000Hz was taken as the general hearing aid latency. In a few cases (subjects S03, S04, and S06), the hearing aid latency was taken from the device delay list provided by MED-EL[1]. The subject-specific hearing aid latencies are shown in Table 1. The hearing aid latency can be compensated for in the MED-EL CI fitting software Maestro starting from Version 9 for processors Sonnet 2 and Rondo 3 or newer.

Table 1: Subject-specific data of bimodal CI-listeners. Estimated hearing aid (HA) latency was generally measured with a hearing aid analyzer unit (median latency in the frequency range between 500 and 4000Hz). Latencies marked with an asterisk were taken from the 'device delay list' provided by MED-EL. The everyday hearing aid latency compensation refers to the programmed hearing aid latency used in the subjects' personal CI processors. "None" indicates that no HA latency compensation was used. Speech perception in quiet was measured at 65 dB SPL with a monosyllabic word test (Freiburg Monosyllables Test).

| ID | Age (years) | Personal processor type | Implant type | De-activated electrodes | Duration of implant use (years) | Estimated HA latency (ms) | Everyday HA latency compensation (ms) | Speech perception in quiet (% correct) | |
|---|---|---|---|---|---|---|---|---|---|
| | | | | | | | | CI | HA |
| **S01** | 58 | Sonnet | Flex26 | EL12 | 4.4 | 8.2 | None | 70 | 65 |
| **S02** | 70 | Sonnet2 | FlexSoft | None | 0.4 | 7.2 | None | 75 | 25 |
| **S03** | 64 | Sonnet | FlexSoft | None | 6.2 | 5.2* | None | 80 | 85 |
| **S04** | 86 | Sonnet2 | FlexSoft | EL9-12 | 1.0 | 9.3* | None | 50 | 70 |
| **S05** | 66 | Sonnet2 | FlexSoft | EL11-12 | 8.9 | 8.3 | None | 65 | 65 |
| **S06** | 60 | Sonnet2 | FlexSoft | None | 1.3 | 6.9* | None | 80 | 100 |
| **S08** | 60 | Rondo3 | FlexSoft | None | 2.1 | 2.5 | 2.6 | 80 | 90 |
| **S09** | 69 | Sonnet2 | Flex24 | None | 6.8 | 8.3 | None | 75 | 35 |
| **S10** | 73 | Sonnet2 | FlexSoft | None | 2.8 | 8.2 | 8.7 | 55 | 50 |
| **S11** | 59 | Rondo3 | FlexSoft | EL12 | 0.8 | 6.3 | 6.3 | 65 | 90 |
| **S12** | 70 | Sonnet2 | FlexSoft | none | 1.0 | 6.1 | 6.0 | 55 | 50 |

---

[1] https://www.medel.pro/online-resources/timing-settings-for-hearing-aids

## Participants

Participants were recruited from the clinical CI-center and all gave written informed consent. The study was approved by the local ethics committee (2020-022). The experiments were conducted in a sound-treated room at the CI-center.

Eleven bimodal MED-EL CI listeners participated in the experiment. Their age ranged from 58 to 86 years. The listeners were fitted with a hearing aid contralaterally to the CI. Both devices were used daily. Unaided pure-tone averaged thresholds at 0.5, 1, 2, and 4 kHz (PTA4) on the hearing aid side ranged from 44 to 80 dB HL. Note that S11 had no measurable thresholds above 2 kHz. Therefore, a threshold of 120 dB HL was used for the PTA4 calculation at 4 kHz in this case. CI thresholds and both aided and unaided thresholds on the hearing aid side are shown in Figure 1.

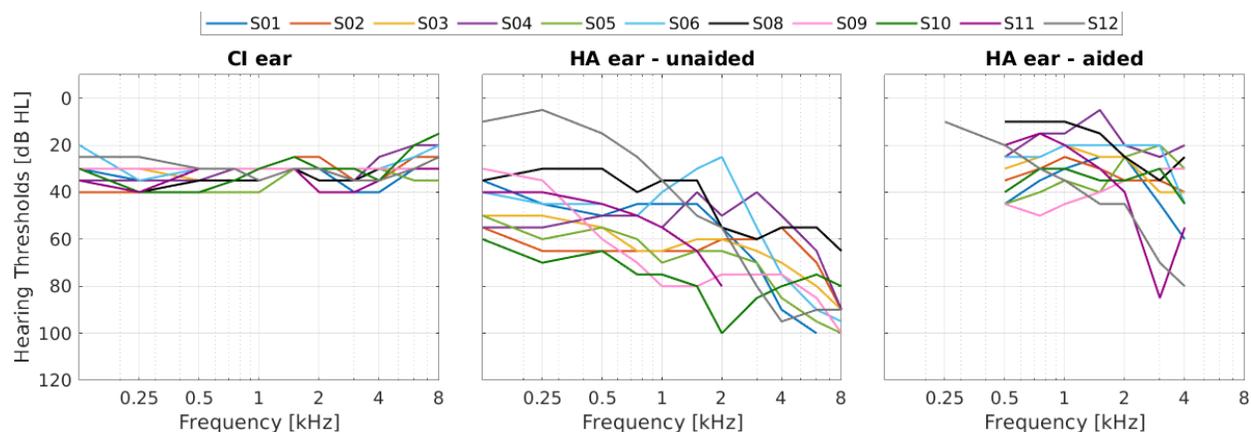

*Figure 1: Individual hearing thresholds with the cochlear implant (CI, left panel), unaided pure-tone thresholds contra-laterally to the implanted ear (middle panel) and aided thresholds with the hearing aid (HA, right panel) of all participants.*

Participants had at least four months of CI experience and achieved 50% or more correct responses with the Freiburg Monosyllables Test (speech perception in quiet) at a level of 65 dB SPL with each device, except for subject S02 and subject S09 who only achieved 25% and 35% correct with the hearing aid, respectively. Four CI users were already fitted with a hearing aid latency compensation $\tau_{add} = \hat{\tau}_{HA}$ on their personal processor. An overview of the subject-specific information is given in Table 1.

## Experimental Setup

Sound source localization measurements were performed in a sound-treated room. The measurement setup consisted of 13 loudspeakers (8030C studio monitors, Genelec, Finland) placed in a semicircle with a radius of 1.35 m at ear height and 15° apart in the frontal azimuthal half-plane. Each stimulus consisted of three 70-ms long Gaussian broadband white noise bursts, each gated with a 5-ms Hanning window, separated by two 30-ms long silent intervals (similar to Angermeier et al., 2021). Each localization measurement consisted of 55 stimulus presentations (5 from each of the 11 most central loudspeakers). The presentations were randomized. Stimuli were presented at a level of 65 dB SPL. Random level roving of ±5dB was applied. Calibration was performed with a hand-held level meter (PCE-322A, PCE-Deutschland GmbH, Germany) at the subject's head position.

All listeners used the same CI processor (MED-EL Sonnet2) during the measurements, which was programmed with the most frequently used individual map of each subject. All listeners, except for subject S04, used the FS4-processing strategy with four fine structure channels in their individual maps. Subject S04 used only three fine structure channels. During testing, the adaptive intelligence program was disabled, and the microphone setting was set to "natural". On the acoustic side, the listeners used their individual hearing aid with their everyday settings. Subject S11 had a slightly louder sound impression with the CI processor used for testing than with the personal processor, a Rondo3. Therefore, the CI-volume was reduced one step in the volume settings, which corresponded to a volume reduction of 4%. All other subjects performed the experiments at their average CI volume setting.

The global hearing aid latency compensation setting in the CI $\tau_{add}$ was systematically varied for each measurement using the MED-EL fitting software Maestro 9. In general, six different latencies between 1.5 and 10 ms were applied (1.5ms, 2/3ms, 4/4.5ms, 6ms, 8ms, and 10 ms). Note that the software requires a minimal CI delay setting of 1.5 ms.[2] If the everyday hearing aid compensation setting of a subject was close to one of the experimental latencies mentioned above, the everyday hearing aid compensation setting was used instead (e.g., subject S11 performed the localization experiment with a latency of 6.3 ms instead of 6 ms). In the case of subject S10 an additional measurement with an additional latency of 12 ms was conducted, since the hearing aid latency measurement using the hearing aid analyzer unit suggested a substantially higher latency in the lower frequencies (median of 11.3ms in the frequency range between 500 and 1000Hz) compared to the higher frequencies (median of 8.1ms). The first localization measurement was performed with the participants' everyday hearing aid compensation setting (see Table 1 for individually used hearing aid compensation settings). The following localization measurements were conducted with randomized latency settings. In addition to the measurements with systematically varying latency settings, a retest measurement was performed at the end of the measurement session. Eight of the eleven subjects performed the retest, which was performed with the same latency setting as used for the first measurement. Four listeners did not perform all six planned latency measurements and retest due to technical problems (S01), fatigue (S04, S09) and pure monaural stimulus perception (S12)

---

[2] The authors only became aware of the minimum latency of 1.5 ms to be set during the measurements. Therefore, the planned latencies were slightly adjusted after subject 08: 3 ms instead of 2 and 4.5 ms instead of 4.

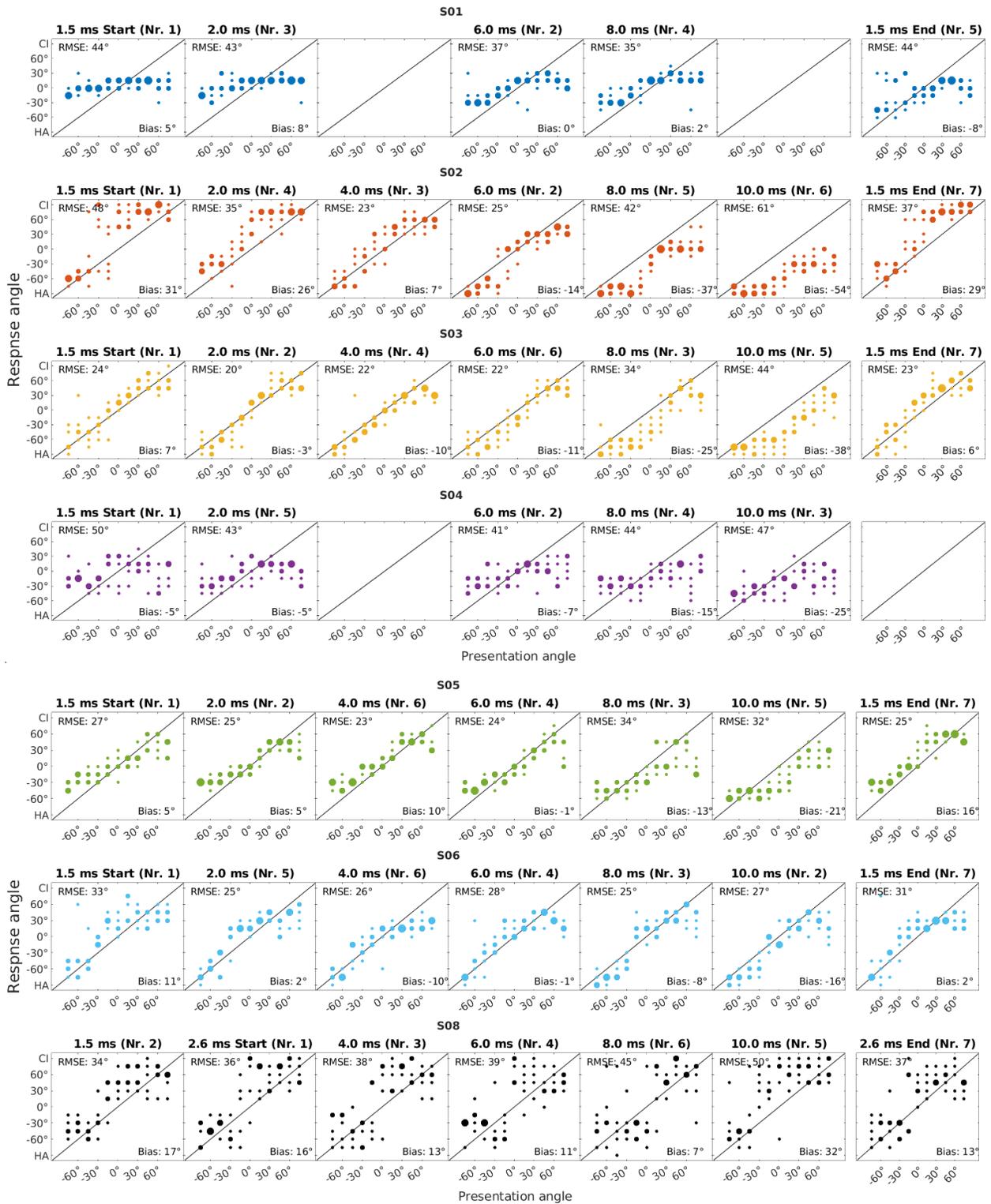

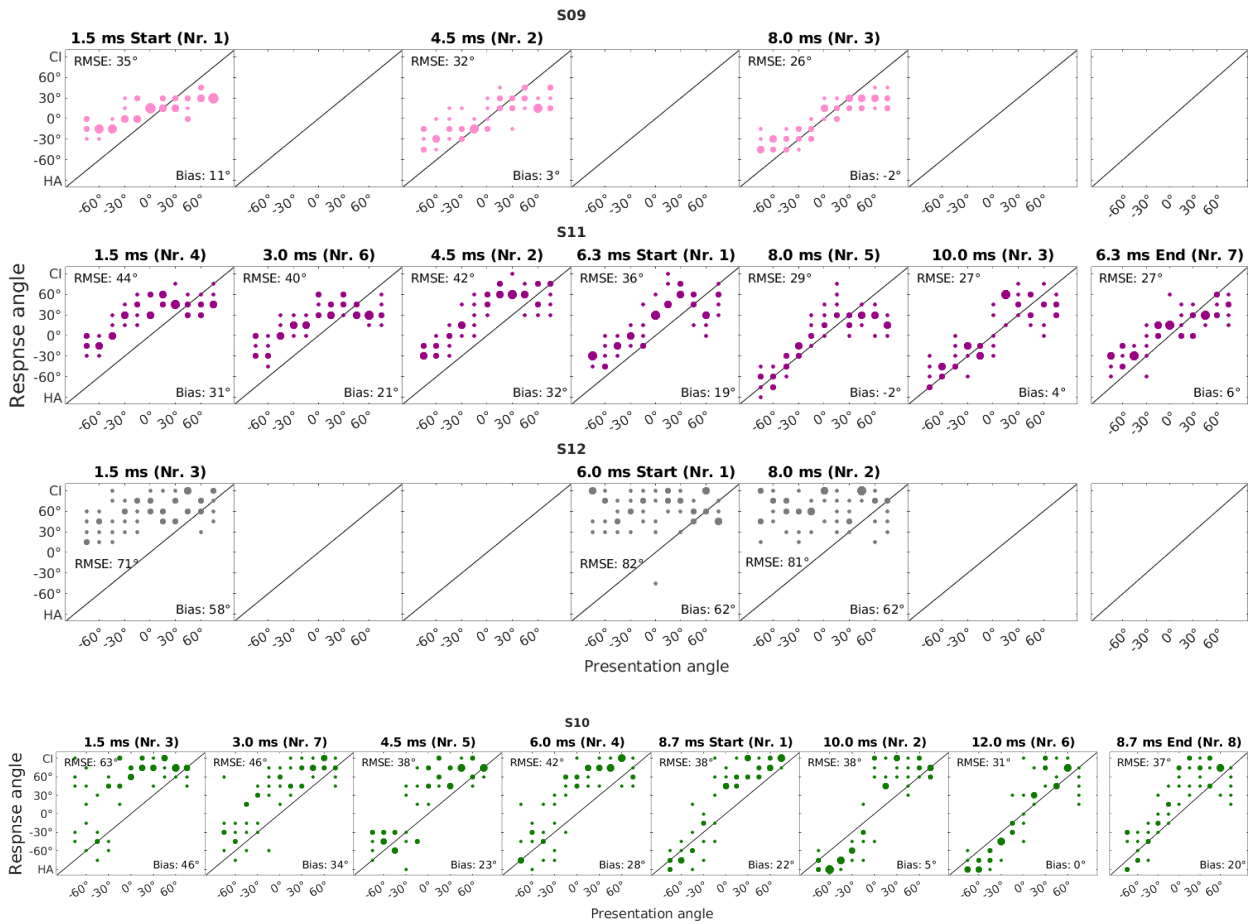

*Figure 2: Bubble plots indicating the response angle over presentation angle for each subject and each of the additional CI latencies measured. Data was mirrored as needed, such that the hearing aid (HA) side is always at -90° and the CI side at +90°. The overall root mean square error (RMSE) and bias is shown in each subplot at the top left and bottom right corner, respectively. The number on top of each plot indicates the measurement order for the respective subject.*

# Results

On average, the root mean square error (RMSE) of the localization with the clinical settings was 39°, compared to 40° for Angermeier et al. (2023). When adjusting the latency settings to minimize the estimated latency mismatch, the mean RMSE improved marginally to 37°, compared to 32° for Angermeier et al. (2023). The mean of the lowest RMSE across all tested latencies was 32°.

Figure 2 shows the responded angle (y-axis) over presentation angle (x-axis), for each subject and each of the latencies added in the CI settings. Negative angles refer to the side of the hearing aid and positive angles to the side of the CI. Each dot indicates a response, and larger dots are used to illustrate the same response for multiple repetitions of a presentation angle. If the active speaker was correctly identified the response would lie on the diagonal, which is illustrated by a line. At the top left of each panel, the RMSE for the current condition is given, and on the bottom right, the bias is depicted.

In cases with little or no added latency we expected a bias towards the CI. With increasing CI latency, we expected a shift in localization performance toward the center, combined with a decrease in RMSE, up to a subject-specific point. As latency was further increased, we expected an increasing bias towards the hearing aid side and again an increase in RMSE, since at some point the hearing aid side has a lower latency compared to the CI side. This pattern was observed in subjects S02, S03, S05 and S06. S03 had an overall lower bias compared to S02 and especially a lower bias towards the CI side. A similar pattern was shown by S10, and S11, although with more noisy responses. Both S05 and S11 tended not to use the full range of possible responses.

Subject S01 tended to localize everything towards the midline with very few responses to both sides. This became slightly less pronounced with increasing latency, but also with practice, as shown in the retest at the end. Similar responses were shown by S04 and S09, with a very noisy response pattern for S04, which did not become much clearer as the added latency was increased.

In contrast to the patterns above, S08 localized everything more to the sides. In the bubble plots themselves, it is difficult to see the influence of added latency, but when looking at the RMSE and bias a systematic effect can be seen. Subject S12 was completely unable to perform the task, as every stimulus was recognized from the CI side, and even there the loudspeaker selection was apparently random.

The first step in the data analysis was to evaluate the RMSE and bias for each condition and subject over the estimated latency difference ($\Delta\hat{\tau}$, Figure 3). A value of zero indicates an equalized latency between CI and hearing aid according to the estimated hearing aid latency. A negative estimated latency difference indicates that the hearing aid is faster than the CI, and thus the CI is faster for positive values. To reduce the influence of the measurement noise, Figure 3 also shows quadratic fits for the RMSE and linear fits for the bias in the dashed lines in addition to the measured values. For the quadratic fits, we have constrained the x-axis shift to the range of the measured values.

We divided the subjects' results into three groups based on their fitted RMSE and bias patterns and plotted them separately. Group 1 (first column in Figure 3) had the lowest fitted RMSE and the lowest fitted absolute bias for a positive estimated latency difference (or faster CI), implying that the estimated hearing aid latency was too long to achieve minimal RMSE and bias. This is the opposite to what Angermeier et al. (2021) reported. For all subjects in group 1, the RMSE shows the U-like shape we expected, with increasing error values as the estimated latency difference moves away from the optimum. This optimum, however, is not a clear minimal value, but multiple latency differences seem to result in a similar error.

Group 2 (second column in Figure 3) had the lowest RMSE and absolute bias for negative estimated latency differences (or faster hearing aid), implying that the estimated hearing aid latency was too short to achieve minimal RMSE and bias, as in Angermeier et al. (2021). The RMSE of the subjects in group 2 just decreased with a higher added latency, and fitting a parabola in this data might seem a bit arbitrary. However, it is reasonable to assume that a further increase in the additional latency would cause the RMSE to increase again, as for group 1.

The bias values for both the groups 1 and 2 had a general tendency to move from the CI side (positive estimated difference) to the hearing aid side as the additional latency increased. For all subjects in these two groups, the bias was close to zero when the RMSE was minimal.

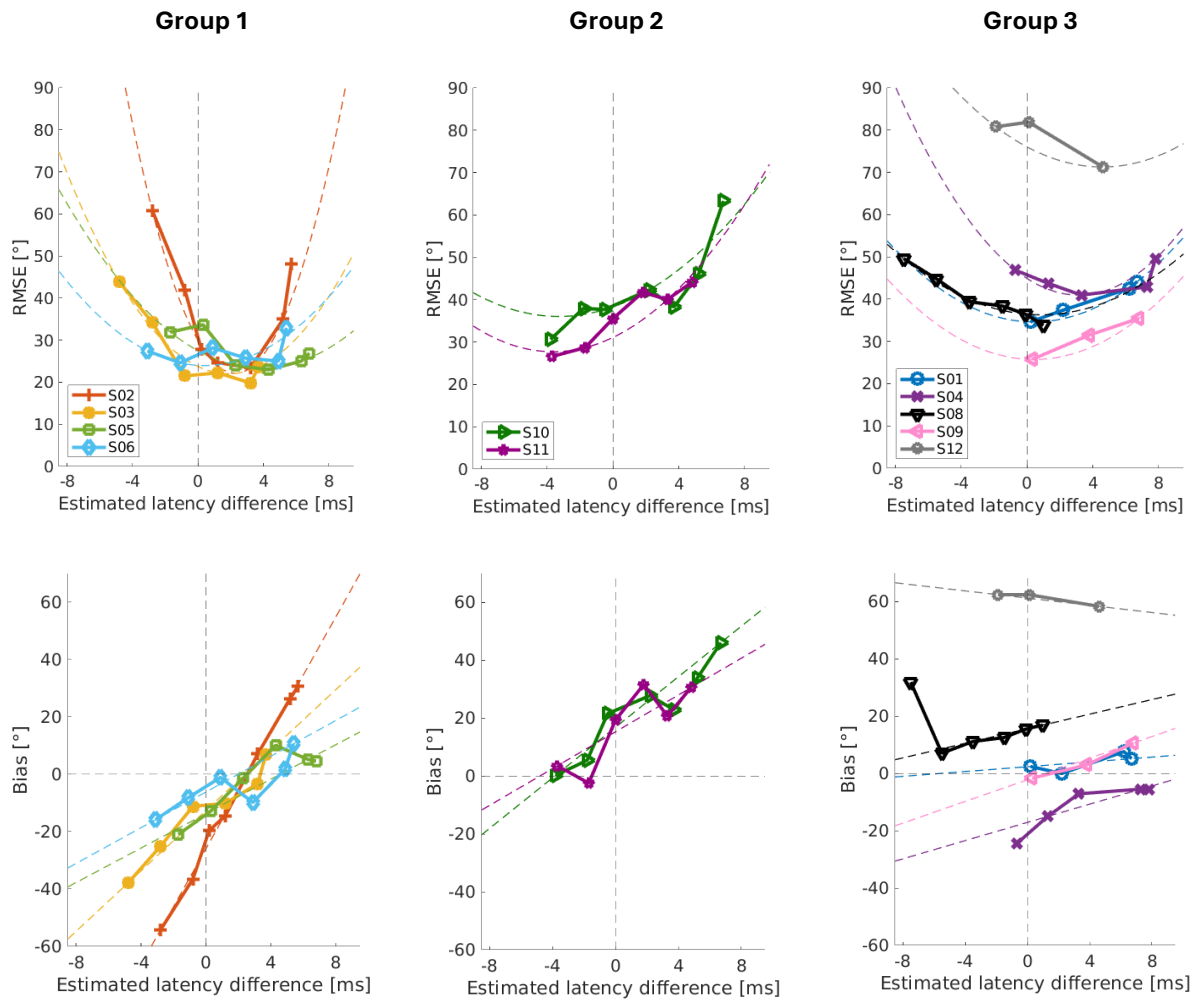

*Figure 3: The upper row shows the root mean square error (RMSE) over the estimated latency difference for the three subject groups. Additionally, to the data, a squared fit for each subject is shown with dashed lines. The lower row shows the bias over latency difference for the same three groups. The bias values have been fitted with a linear function, also shown with dashed lines. The first group contains the subjects that have the best latency difference slightly on the CI side. The second group contains those subjects with the best latency on the hearing aid side. Finally, the third group contains the subjects who had trouble performing the task and could therefore not be classified in group 1 or 2.*

Finally, for subjects in group 3 (third column in Figure 3), there was either not enough data or the data was not clear enough to decide on one of the previous groups. This includes those subjects who had a strong tendency to localize either to the front or to one or both sides. Regardless of the location of the lowest RMSE and bias, it can be seen that the latency difference has a systematic influence on the localization performance as described by RMSE and bias for all subjects, except for S12, who was not able to localize at all. All subjects in group 3 had some problems with localization, which is reflected in the higher RMSE values. Nevertheless, the RMSE values changed systematically when the latency was changed, except for S12. This shows that even subjects with limited binaural fusion can benefit from adjusting the latency. However, the bias values in group 3 are also less clear, and for S08 and S12 a large bias towards the CI side remains even for large values of $\tau_{add}$.

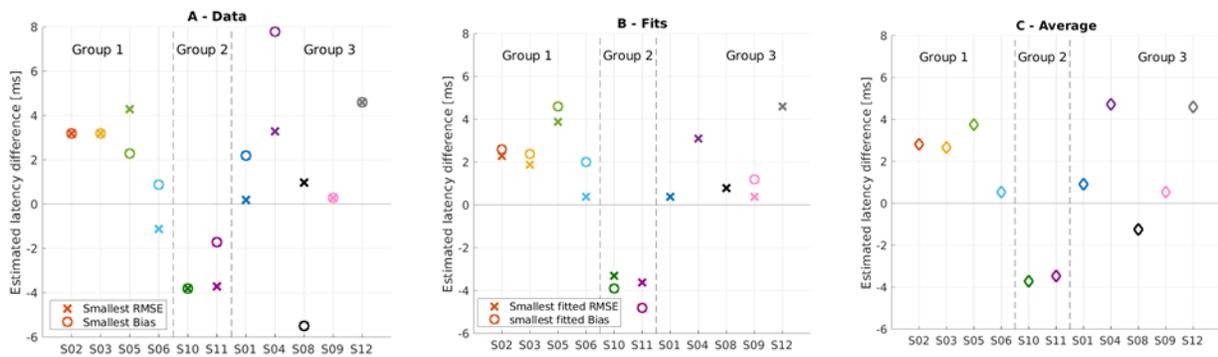

*Figure 4: The plot in panel A shows the best estimated latency difference based on the values leading to the lowest RMSE and the lowest bias, as obtained the measured data. In B the estimated latency difference of the lowest RMSE and bias is shown, as obtained from the fits of the RMSE and bias data. C shows the average estimated latency difference of the lowest RMSE and bias of the values shown in panels A and B.*

Figure 4 compares the estimated latency differences resulting in the lowest RMSE and lowest bias in different ways. Panel A shows the estimated latency difference of the lowest measured RMSE and bias. For five of the eleven subjects, the estimated latency difference based on the lowest measured bias and the lowest measured RMSE led to the same results. For the other subjects, there seemed to be a gap between the two values. However, this may be partly due to the very discrete step size of 1.5 to 2 ms used in this experiment. Only for S04 and S08 the best estimated latency differences based on bias and RMSE were more than one step apart. Panel B in Figure 4 shows the estimated latency difference of the lowest RMSE resulting from the quadratic fit and of the bias of 0° resulting from the linear fit. Here, both estimated latency differences are very close together with a maximal distance of 1.6 ms for groups 1 and 2. For group 3, the best estimation of latency difference based on the bias was difficult because not all linear fit functions crossed zero close to the measured range. In these cases (S01, S04, S08, S12), the best latency difference is not shown.

The best latency difference based directly on the measured data and based on the data fits differed by up to 2 ms in most subjects. In one case (bias of S11) it differed even more due to the highly non-monotonic data. These numbers match with our visual inspection, that best latency estimates have an uncertainty of 1-2 ms.

To obtain a single value for the best estimated latency difference, we have taken the mean value of the best estimated latency difference based on the RMSE and bias of the measured and fitted values (Figure 4, panel C). It is interesting to note that only for three subjects the estimated latency difference was close to 0, which we define as the interval between -1 ms and 1 ms. This indicates that matching the latencies between the two ears is not a straightforward task, especially if only a frequency independent additional latency is available.

Figure 4C also clearly shows that for group 1 the best localization performance was obtained for a positive estimated latency difference (faster CI), even though for S06 the best latency was close to 0 and, if based only on the RMSE data, the best latency was negative. For group 2, the best localization performance was obtained for a negative latency difference (faster hearing aid). Overall, most of our subjects seemed to localize best for Δτ values of 0 to 5 ms, i.e., when $\tau_e \leq \tau_a$, or equivalently when $\tau_{add} \leq \tau_{HA}$.

In Figure 5, the RMSE values of the subjects for different conditions were compared pairwise. This was done by plotting the values in one condition against the values in the second condition. In this way, data points above the diagonal indicate larger RMSE in condition marked on the y-axis as compared to the condition on the x-axis. A greater distance from the diagonal indicates a greater difference between the two conditions.

To evaluate the effect of training, we measured the same condition at the beginning and at the end (retest) of the measurement sessions for eight of the eleven subjects. We used the latency settings that the subjects wore every day. The measured RMSE values of the start and end conditions can be seen in Figure 5 panel A. When comparing the values, only a small effect of training is visible (mean RMSE improvement: 3°) and most subjects were close to the diagonal (Fig. 5A). However, as only one subject was slightly worse in the end condition and all other subjects improved their performance, an exact two-tailed Wilcoxon signed rank test showed a statistically significant difference (p=0.02).

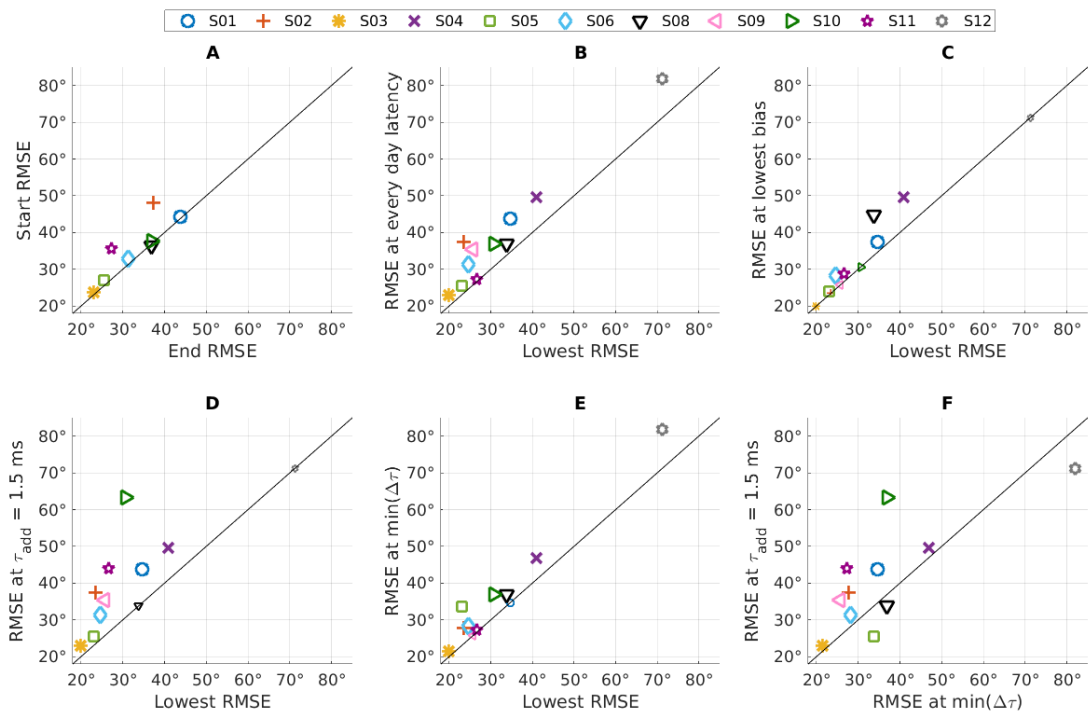

Figure 5: Pairwise comparison of the root mean square error (RMSE) for multiple conditions. The symbols indicate the data points of the individual subjects. Values above the diagonal indicate a lower RMSE value for the condition on the x-axis compared to the conditions on the y-axis. A higher distance from the diagonal rates to a higher difference between the two conditions. If the values are the same for both conditions, they lie on the diagonal and are shown with small symbols.

In the next step, we compared the RMSE measured with the everyday latency setting with the lowest measured RMSE of the respective participant obtained at any latency (Figure 5, panel B). For the eight subjects where a test-retest was done, the value of the retest measurement was used. The comparison showed a small improvement in the RMSE for every single subject, independent of the localization ability. The lowest RMSE is, per definition, always lower or equal to the value it is compared to. Therefore, a test for statistical difference would not be meaningful, as the difference would always be significant if not all values are equal. On average, the RMSE was reduced by 7° (17%) from 39° to 32°.

Following up on the comparison of the best latency estimated based on the lowest RMSE with the best latency estimated based on lowest bias (Fig. 4), we now compare the corresponding RMSE values. For this, we plotted the RMSE at the lowest bias against the overall lowest RMSE in Figure 5, panel C. This comparison shows that even if the best latencies estimated based on bias or RMSE differed, the difference in RMSE for both conditions was on average only 3° (lowest RMSE: 32°, RMSE at lowest bias: 35°).

In the next step, we compared the RMSE measured with the lowest possible additional latency with the overall lowest measured RMSE (Figure 5, panel D). This central comparison demonstrates the total potential of latency optimization. With the present version of the programming software, the lowest additional latency is $\tau_{add} = 1.5$ ms. Only for S08 and the poorly localizing S12, the RMSE with $\tau_{add} = 1.5$ ms resulted in the lowest RMSE. The average improvement from the RMSE of the lowest additional latency to the lowest overall RMSE was 9° from 42° to 32°.

The simplest way to compensate for a latency difference in the clinical routine would be to set the additional latency equal to the hearing aid latency, as it was done by Angermeier et al. (2023). This results in an estimated latency difference $\Delta\tau = 0$. Since we did not measure this value directly, we used the lowest estimated latency difference $min(\Delta\tau)$ instead. This value was between ±1 ms for all subjects. In Figure 5, panel E, we compare the RMSE at $min(\Delta\tau)$ with the overall lowest RMSE. The average improvement for all subjects from the RMSE at $min(\Delta\tau)$ to the overall lowest RMSE value was 4° from 37° to 32°.

Finally, we compared the RMSE of the lowest estimated latency difference ($min(\Delta\tau)$) with the RMSE obtained when no additional latency was added ($\tau_{add} = 1.5$ ms) (Figure 5, panel F). This shows that for most subjects a simple device latency-based compensation was beneficial, but for three of the eleven subjects the performance decreased. On average the improvement in RMSE was 5°, but this was not statistically significant (exact two-tailed Wilcoxon signed rank test, p = 0.21).

## Discussion

In this study, we investigated the localization abilities of bimodal CI users with multiple latency differences between the CI and the hearing aid side. We have found that adjusting the additional CI latency immediately improved the localization ability of our bimodal CI subjects, which is consistent with the studies of Angermeier et al. (2021, 2023).

The estimated latency difference that produced the best localization performance, both in terms of lowest RMSE and lowest absolute bias, was for most subjects not close to 0 ms estimated latency difference. There are several possible reasons for this. The first is that the latencies in both devices are frequency dependent. A frequency-independent compensation of the latency difference is certainly suboptimal, and the best value is likely to depend on both the dominant frequency band and on the frequency dependence of the hearing aid latency, which differs across manufacturers.

A second reason could be an adaptation to the latency difference. Although adaptation to the latency difference seems unlikely and the acute improvement in localization performance in our study and in the studies by Angermeier et al. (2021, 2023) argues against it, partial adaptation could play a role. None of the subjects in group 1 had a compensation for the latency difference in their clinical map, although all subjects had an estimated latency difference of at least 6 ms (Table 2). Looking at the RMSE over the estimated latency difference (Figure 3), it can be seen that all four subjects in this group benefit most from a CI faster than the hearing aid side. The theory of partial adaptation to a latency difference is also supported by the fact that both subjects in group 2 have a latency compensation in their clinical map and benefit the most from a CI slower than the hearing aid side. Interestingly, this is not observed in the data of Angermeier et al. (2021). In their study, most subjects preferred a slightly slower CI side even though none of them had a compensation for the latency mismatch. However, only additional CI latencies within ±1 ms of the estimated hearing aid latency were used by them.

Another reason for the best localization performance at non-zero latency difference could be an interaction between level mismatch and latency difference, as also suggested by Pieper et al. (2022). A level mismatch between the two ears causes an offset in the interaural level difference (ILD), which itself causes a localization bias. It is possible that this bias is partially compensated for by an opposite offset in interaural time difference (ITD), i.e., a latency difference. This could also be the reason for the partial adaptation to the latency difference mentioned above. If the fitting procedures for both the hearing aid and the CI sides aim to achieve a balanced percept, this may result in partial compensation of the latency mismatch by a level difference. This partial compensation could then appear as an adaptation to the latency difference.

An interaction between the level mismatch and the latency difference can also be seen as an opportunity. A central goal of bimodal fitting is to achieve the best possible speech understanding for sounds from all directions. This might require an independent fitting of levels and gains on the two devices which can result in an interaural level mismatch. In other words, even when latency is compensated, there may be a level bias in sound localization, and compensating the level bias may reduce speech intelligibility. Our results suggest that the latency difference can be viewed not only as a problem but also as a degree of freedom, that can be used to partially compensate for a localization bias without compromising speech understanding in either ear.

If we compare the RMSE when no additional latency is used with the RMSE for a minimal latency difference, we find an improvement in the performance for eight of our eleven subjects, similar to Angermeier et al. (2023). However, this reduction was not significant at the a group level in either our study or in Angermeier et al. (2023). In contrast, Angermeier et al. (2021) found a significant improvement in localization performance. A possible reason for this difference is that the current clinical software, which allows to add an additional latency, does not allow an additional latency below 1.5 ms. This means that the baselines used by Angermeier et al. (2023) and in our study differ from those at Angermeier et al. (2021) by an additional latency of 1.5 ms. Comparing the baseline performances, it can be seen that the initial RMSE also improved from 53° in Angermeier et al. (2021) to 40° in Angermeier et al. (2023). As such the new additional latency in MED-EL devices appears to be beneficial for bimodal patients that do not get a latency fitting, but potentially detrimental to some single-sided deaf CI users without a hearing aid.

Looking at the results of the individual subjects, it is interesting to note that both S08 and S12 had the lowest RMSE at an additional latency of 1.5 ms (Figure 5, panel D). The audiograms (Figure 1, panel B) show that these two subjects still have good hearing in the low frequency range, which might allow for direct sound perception at these frequencies. If the low frequencies dominate the sound localization abilities, the best latency difference needs to be matched at the low frequencies. These two subjects illustrate once again that matching the latencies is not a straightforward task. It is highly dependent on the individual patient and not only on the hearing aid worn.

Both S08 and S12 are also two of the worst performing subjects. For S08, the bubble plots in Figure 2 show a pattern where everything is localized to the sides. This indicates an unfused sound perception. S12 perceived all sound sources as coming from the CI side. Although the RMSE was slightly lower for an additional latency of 1.5 ms, it was still much worse than the RMSE of any other subject in this study. Looking again at the audiogram, we can see that in addition to good low frequency hearing, S12 also has very bad high frequency hearing on the acoustic side. This might contribute to the advantage of the CI ear, especially since our stimuli were white noise bursts, which contain a higher amount of high frequency energy compared to most real-world stimuli.

Another subject with a rather poor localization performance is S04. Here, the non-monotonic change of the response with changing additional latency (see Figure 4) is particularly interesting. Multiple possible reasons arise when looking at the meta data. The first one is that S04 is by far the oldest of our subjects and at 86 years of age may have had some difficulty in performing this task. Additionally, four electrodes of the CI of S04 are deactivated. This could be related to a shallow insertion.

In the Introduction we hypothesized, in agreement with Pieper et al. (2022), that it should be possible to improve the localization abilities of bimodal CI users to a performance comparable to that of bilateral CI users. Therefore, we now compare the best possible localization results of the eleven subjects in our study with the localization performance of the study by Dorman et al. (2016). It can be seen that just by finding the optimal latency difference for our subjects, a mean RMSE of 32° (median RMSE: 27°) can be achieved. This performance is very comparable to that of the bilateral CI users in the study by Dorman et al. (2016), who achieved reached a mean RMSE of 29° and a median of 27°. This shows that bimodal CI users are able to localize similarly to bilateral CI users when their devices latencies are optimally fitted.

The stimuli used in this study were Gaussian white noise bursts, as in (Angermeier et al., 2021; 2023). They were used because they facilitate the measurement of temporal localization abilities. However, stimuli with such a broad spectrum and sharp onsets are rare in real-world stimuli and may have exaggerated the observed effect. More natural stimuli with shallower envelopes have a longer envelope coherence length and the latency between the stimuli is less critical, as discussed by Pieper et al. (2022). It is therefore conceivable that a smaller benefit of the best latency would have been observed with more natural stimuli. It is also possible that stimuli with different spectral content lead to a different best latency difference, due to the frequency dependence of the latencies in the CI and the acoustic hearing side. For clinical applications, the use of a more natural stimulus, such as speech, may be beneficial.

We expect that an improved localization performance will improve hearing in everyday listening situations. Binaural fusion is also facilitated by interaurally coherent input (Brown & Tollin, 2021). As the latest generation of MED-EL CIs allows to set an additional CI latency, it is recommended to take advantage of this opportunity to improve the hearing of the patients. However, it is not feasible to perform the presented localization test during the clinical routine. As the difference between the lowest RMSE and the RMSE for the best bias (Figure 5, bottom left) is small, we suggest that the bias can be used as a basis for a clinical latency fitting routine. The bias could be estimated during the fitting by asking the patient from which direction (left, front, right) a sound was heard. The latency is then adjusted until a sound from the front is perceived as coming from the front. If patients are unable to perform this task, estimating the difference based on the hearing aid latency difference is also likely to improve the localization abilities.

## Conclusion

In our study, we measured the localization performance of eleven bimodal CI subjects for multiple latency differences between the two ears. We showed that adjusting the latency difference immediately improved localization performance for all subjects. Interestingly, the latency difference that produced the best localization performance was rarely 0 ms. Most subjects preferred a CI that was 2-5 ms faster than the calculated match, but two subjects performed best with a CI that was approximately 4 ms slower than calculated. We conclude that adjusting the latency difference is beneficial for binaural hearing. However, minimizing the latency difference is often not ideal and a dedicated test to find the optimal latency difference for the patient may be necessary. With the best possible latency difference for each subject, the average RMSE in our study was comparable to that of bilateral CI users in the study by Dorman et al. (2016). This shows that the bimodal CI users can achieve the same localization performance as bilateral CI users. Our results indicate that the latency difference can be considered as a degree of freedom, that allows to compensate for a localization bias due to a level mismatch between the two ears. We recommend that adjusting the latency difference should be part of the clinical fitting routine.

## Funding

The authors disclose receipt of the following financial support for the research, authorship, and/or publication of this article. This work was supported by the European Research

Council (ERC) under the European Union's Horizon 2020 Research and Innovation
Programme (ERC Starting Grant to Mathias Dietz; grant number 716800).

## Declaration of Conflicting Interests

The authors declare no potential conflicts of interest with respect to the research, authorship and/or publication of this article.

## Data availability

The data is freely available at doi.org/10.5281/zenodo.15302629